\documentclass[pra,twocolumn,superscriptaddress,longbibliography]{revtex4-1}
\usepackage{amssymb,amsmath,amsfonts,bm}
\usepackage{graphicx}
\usepackage{epstopdf}
\usepackage{times}
\usepackage{float}
\usepackage{lipsum}
\usepackage{color}

\usepackage{hyperref}
\hypersetup{  colorlinks=true, linkcolor=blue, citecolor=red, urlcolor=blue  }

\usepackage{physics}

\begin{document}

\title{Nonadiabatic Noncyclic Geometric Quantum Computation in Rydberg atoms}

\author{Bao-Jie Liu}

\affiliation{Department of Physics, Harbin Institute of Technology, Harbin 150001, China}
\affiliation{Institute for Quantum Science and Engineering, and Department of Physics,
Southern University of Science and Technology, Shenzhen 518055, China}

\author{Shi-Lei Su} \email{slsu@zzu.edu.cn}
\affiliation{School of Physics, Zhengzhou University, Zhengzhou 450001, China}

\author{Man-Hong Yung}  \email{yung@sustech.edu.cn}

\affiliation{Institute for Quantum Science and Engineering, and Department of Physics,
Southern University of Science and Technology, Shenzhen 518055, China}

\affiliation{Guangdong Provincial Key Laboratory of Quantum Science and Engineering, Southern University of Science and Technology, Shenzhen 518055, China}

\affiliation{Shenzhen Key Laboratory of Quantum Science and Engineering, Shenzhen 518055, China}

\date{\today}

\begin{abstract}
Nonadiabatic geometric quantum computation (NGQC) has been developed to realize fast and robust geometric gate. However, the conventiona NGQC is that all of the gates are performed with exactly the same amount of time, whether the geometric rotation angle is large or small, due to the limitation of cyclic condition. Here, we propose an unconventional scheme, called nonadiabatic noncyclic geometric quantum computation~(NNGQC), that arbitrary single- and two-qubit geometric gate can be constructed via noncyclic non-Abelian geometric phase. Consequently, this scheme makes it possible to accelerate the implemented geometric gates against the effects from the environmental decoherence. 
Furthermore, this extensible scheme can be applied in various quantum platforms, such as superconducting qubit and Rydberg atoms. Specifically, for single-qubit gate, we make simulations with practical parameters in neutral atom system to show the robustness of NNGQC and also compare with NGQC using the recent experimental parameters to show that the NNGQC can significantly suppress the decoherence error. In addition, we also demonstrate that nontrivial two-qubit geometric gate can be realized via unconventional Rydberg blockade regime within current experimental technologies. Therefore, our scheme provides a promising way for fast and robust neutral-atom-based quantum computation.
\end{abstract}


\maketitle

\section{Introduction}
Neutral atoms that interacting via dipole-dipole interactions have became a potential platform for quantum computation~\cite{Brennen1999,Weiss2017}. Rydberg atoms are one kind of neutral atoms that are excited to high-lying Rydberg states~\cite{Gallagher1994}, which would exhibit strong Rydberg dipole-dipole interaction when the inter-atomic distance is not very large. And the Rydberg-Rydberg interaction~(RRI)~have been studied for construction of quantum logic gates~\cite{Jaksch2000,Lukin2001,SaffmanRMP}. By using microwave transitions, single-qubit Rabi oscillation of neutral atoms have been well studied experimentally~\cite{Schrader2004}. Besides, high fidelity single-qubit quantum logic gates~\cite{Olmschenk2010,xia2015L,Wang2015,Lee2013} and quantum controls~\cite{Smith2013} have also been demonstrated in neutral atoms. Through the laser-induced transitions from ground state to Rydberg state, many two- and multiple-qubit gates in neutral atom based on RRI have also been demonstrated in experiments~\cite{Isenhower2010,Wilk2010,Maller,Zeng2017,Picken2018,Levine2018,Levine2019,Graham2019,Is2020}. These experimental studies show the high-fidelity of single-qubit gates and also show how to improve the fidelity of two-qubit gates step by step. On that basis, if one can design single- and two-qubit quantum logic gates that are more robust to systematic fluctuation error and decoherence, it will be beneficial to realize quantum computation in neutral atoms.

Geometric quantum logic gates~\cite{gqc,Zanardi1999} based on adiabatic or non-adiabatic geometric phase~\cite{b2,b3,b1,b4}, which depends only on the global properties of the evolution paths, provides us the possibility for robust quantum computation~\cite{Zhu2005,Berger2013,Chiara,Leek,Filipp,Thomas,Johansson}. In contrast to the earlier adiabatic-process-based 
geometric quantum computation~\cite{Duan2001,lian2005,Jones2000,Huang2019}, non-adiabatic geometric quantum computation (NGQC) and  non-adiabatic holonomic quantum computation (NHQC) based on Abelian~\cite{b5,b6,liang2016,zhao2017,Chen2018,Zhang2020,XXu2018,ZZhao2019} and non-Abelian geometirc phases~\cite{Sjoqvist2012,Xu2012,liu2019,liu2020,xue2017,Zhou2018,Mousolou2017,Song2016NJP,Liu2017PRA,Kang2020} in two- and three-level system, respectively, can intrinsically protect against environment-induced decoherence,
since the the construction times of geometric quantum gates is reduced. The non-adiabatic geometric gates of NGQC and NHQC have been experimentally demonstrated in many systems including superconducting qubit~\cite{Abdumalikov2013,Song2017,stasuper2018,XXu22018,Han2020}, NMR~\cite{Feng2013,li2017,zhu2019,Yingcheng2020},  NV center in diamond~\cite{Zu2014,Arroyo2014,s0,s2}. On the other hand, there are many theoretical proposals to apply geometric quantum computation~\cite{moller2008,zheng2012,beterov2013,wu2017} and NGQC~\cite{zhao2017,zhao2018,kang2018,shen2019,liao2019} in Rydberg atom platform.  To further speed up NGQC scheme, NGQC is incorporated with the time-optimal technology to realize the geometric gate with the minimum gate under the framework of cyclic evolution~\cite{Liu2020arx,Chen2020arx,Chen2020ar}. However, both NGQC and time-optimal NGQC should satisfy the cyclic condition, which leads to neutral-atom-based quantum logic gates being more sensitive to decay and dephasing errors compared to the conventional dynamical counterparts~\cite{Olmschenk2010,xia2015L,Wang2015}. Particularly, the evolution time of NGQC should be exactly the same for all quantum logic gates no matter the geometric rotation angle is large or small.

Here, we propose an new scheme, nonadiabatic noncyclic geometric quantum computation~(NNGQC), that all of single-qubit geometric gate and nontrivial two-qubit can be realized via noncyclic non-Abelian geometric phase in a Rydberg system. Comparing with the conventional Rydberg blockade~\cite{Jaksch2000,Lukin2001,SaffmanRMP}, we consider RRI-induced blockade process seriously by second-order dynamics, which may be more accurate since we do not discard the stark shifts relevant to the ``blockade''. More importantly, our scheme can further reduce the geometric gate time of NGQC~\cite{zhao2017,Chen2018,Zhang2020,XXu2018,ZZhao2019} without the limitation of cyclic condition. Specifically, we found that the certain gate time of NNGQC can be reduced by half compared with NGQC by choosing proper control parameter. The numerically thorough analysis show that, under the same experimental conditions, the decay and dephasing error caused by the environmental noise can significantly be suppressed via our NNGQC rather than the conventional NGQC. Comparing with recently non-cyclic schemes~\cite{Friedenauer2003,Lv2020}, (i) our scheme only needs to adjust the amplitude and phase of microwave field without complicated pulse sequences in a resonant two-level system; (ii) the gate speed of our scheme is faster than Ref.~\cite{Lv2020} with the same Rabi frequency. Furthermore, our NNGQC can also be conveniently applied to other physical platforms such as superconducting qubits~\cite{Abdumalikov2013} and nitrogen-vacancy centers~\cite{Zu2014,Arroyo2014}.

\begin{figure}[tbp]
\centering\includegraphics[width=8cm]{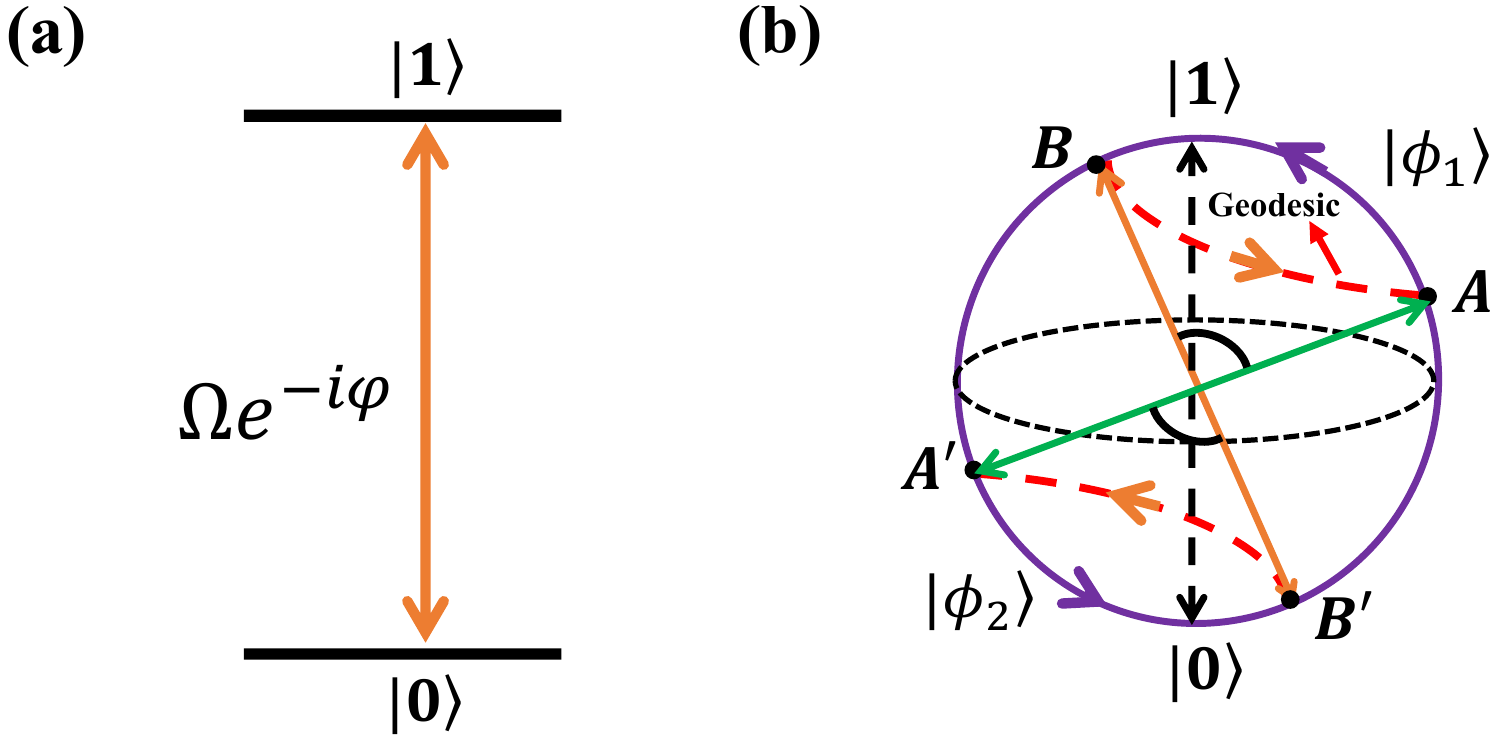}\caption{\label{setup} The illustration of our proposed implementation. (a) The two-level energy structure is resonantly driven by a microwave pulse to realize the transitions of $|0\rangle \leftrightarrow|1\rangle$ with the Rabi frequency $\Omega(t)$ and phase $\varphi(t)$. (b) Conceptual explanation for noncyclic geometric quantum operation. Noncyclic gemetirc phase is given by half the solid angle enclosed by the trajectory $AB (A'B')$ and the geodesic $BA (B'A')$ connecting the initial and final points.}\label{fig1}
\end{figure}

\section{General model}
Here, we consider a $^{133}$Cs with magnetically insensitive ``clock'' states encoding $|0\rangle\equiv|6S_{1/2},~F = 3,~ m_{F}=0\rangle$ and $|1\rangle\equiv|6S_{1/2},~F = 4,~ m_{F}=0\rangle$~\cite{Jaunatphys}, which is resonantly driven by a microwave pulse to realize the transitions of $|0\rangle \leftrightarrow|1\rangle$ with the Rabi frequency $\Omega(t)$ and phase $\varphi(t)$, as shown in Fig. \ref{fig1}(a). The Hamiltonian for this system is (here and after $\hbar \equiv 1$) $H(t)=\frac{\omega_{0}}{2}(|1\rangle\langle 1|-|0\rangle\langle 0|)+\Omega(t)[\cos(\omega_{0}t+\phi(t))|1\rangle\langle 0|+h.c]$, where $\omega_{0}$ denotes the corresponding transition energy.  In the rotating wave approximation and the interaction frame, the system is given by a time-dependent Hamiltonian 
\begin{equation}\label{Original}
H=\frac{1}{2}\left(\begin{array}{cc} 0 & \Omega(t)e^{-i\varphi(t)} \\ \Omega(t)e^{i\varphi(t)}& 0\end{array}\right).
\end{equation}
For a pair of basis vectors $\left\{\left|\psi_{1}(t)\right\rangle,|\psi_{2}(t)\rangle\right\}$ following the Schr\"{o}dinger’s equation as $\left|\psi_{1,2}(t)\right\rangle=\mathcal{T} e^{-i \int_{0}^{t} H\left(t^{\prime}\right) d t^{\prime}}\left|\psi_{1,2}(0)\right\rangle$, the time-evolution operator can be given by $U(t, 0)=\mathcal{T} e^{-i \int_{0}^{t} H\left(t^{\prime}\right) d t^{\prime}}=\left|\psi_{1}(t)\right\rangle\left\langle\psi_{1}(0)\right|+\left|\psi_{2}(t)\right\rangle\left\langle\psi_{2}(0)\right|$. Now, we take a set of auxiliary states $\left|\phi_{1}(t)\right\rangle=\left(\cos \frac{\chi}{2} \mathrm{e}^{-\mathrm{i} \frac{\eta} {2}},\sin\frac{\chi}{2}  \mathrm{e}^{\mathrm{i} \frac{\eta} {2}}\right)^{\text{T}}$ and $\left|\phi_{2}(t)\right\rangle=\left(\sin \frac{\chi}{2} \mathrm{e}^{-\mathrm{i} \frac{\eta} {2}},-\cos\frac{\chi}{2}  \mathrm{e}^{\mathrm{i} \frac{\eta} {2}}\right)^{\text{T}}$, with the boundary conditions $\left|\phi_{m}(0)\right\rangle=\left|\psi_{m}(0)\right\rangle$ at time $t=0$. In this way, $\left|\psi_{m}(t)\right\rangle$ can be expressed $\left|\psi_{m}(t)\right\rangle=\sum_{l} C_{l m}(t)\left|\phi_{l}(t)\right\rangle$, and  the time-evolution operator becomes $U(t,0)=\sum_{l,m} C_{l m}(t)\left|\phi_{m}(t)\right\rangle\left\langle\phi_{m}(0)\right|$. Using the Schr\"{o}dinger’s equation, we obtain the final time evolution operator $U(\tau, 0)=\sum_{l,m=1}^{2}\left(\mathbf{T} \mathrm{e}^{\mathrm{i} \int_{0}^{\tau}({\bf A}(t)+{\bf K}(t)) \mathrm{d} t}\right)_{lm}\left|\phi_{l}(\tau)\right\rangle\left\langle\phi_{m}(0)\right|$, with ${\bf A}_{lm}\equiv i\left\langle\phi_{l}(t)|(d / d t)| \phi_{m}(t)\right\rangle$ being the matrix-valued connection one-form and  ${\bf K}_{l m}(t) \equiv-\left\langle\phi_{l}(t)|H(t)| \phi_{m}(t)\right\rangle$ being dynamical part. If 
a special auxiliary state is chosen to make dynamical part varnished, a noncyclic holonomy matrix can be obtained~\cite{{Kult2006}}.

To realize a noncyclic geometric gate, we choose the auxiliary state $\left|\phi_{m}(t)\right\rangle$ to be proportional to the dynamical states $\left|\psi_{m}(t)\right\rangle$, which satisfies the von Neumann equation~\cite{liu2019}: $\frac{d}{d t} \Pi_{m}(t)=-i\left[H(t), \Pi_{m}(t)\right]$,  where $\Pi_{m}(t)\equiv\left|\phi_{m}(t)\right\rangle\left\langle\phi_{m}(t)\right|$ denotes the projector of the auxiliary basis. Explicitly, we found that they are governed by the following coupled differential equations: 
\begin{equation}\label{DFQ}
\Omega(t)=\frac{\dot{\chi}}{\sin \left(\varphi-\eta\right)}, \quad \varphi(t)=\eta-\arctan \left(\frac{\dot{\chi}}{\dot{\eta}\tan\chi} \right).
\end{equation}
In this way, time-evolution operator becomes: 
\begin{equation}\label{NONGATE}
U(\tau,0)=e^{i\gamma}\left|\phi_{1}(\tau)\right\rangle\langle\phi_{1}(0)|+e^{-i\gamma}| \phi_{2}(\tau)\rangle\left\langle\phi_{2}(0)\right|,
\end{equation}
where $\gamma(\tau)=\int_{0}^{\tau}({\bf A}_{11}+{\bf K}_{11})dt=-\int_{0}^{\tau}({\bf A}_{22}+{\bf K}_{22})dt=\int^{\tau}_{0}\frac{\dot{\eta}}{2\cos{\chi}}dt$ denotes global phase including the  diagonal geometric phase $\gamma_{g}=\int_{0}^{\tau}{\bf A}_{11}dt=\int_{0}^{\tau}\frac{1}{2}\dot{\eta}\cos\chi dt$ and diagonal dynamical phase $\gamma_{d}=\int_{0}^{\tau}{\bf K}_{11}dt=-\int_{0}^{\tau}\Omega\cos\chi\cos (\varphi-\eta) dt$. To make evolution gate in Eq. (\ref{NONGATE}) purely geometric, we set $\varphi-\eta=\pi/2$ for erasing the diagonal dynamical phase. Therefore, the diagonal  geometric phase $\gamma=\int_{\eta(0)}^{\eta(\tau)}\int_{\chi(0)}^{\chi(\tau)}\frac{1}{2}\sin\chi d\chi d\eta=\Omega_{\text{angle}}/2$ is given by half the solid angle enclosed by the trajectory and the geodesic connecting the initial and final points, as shown in Fig. \ref{fig1}(b). Finally, the evolution operator in the basis $\{|0\rangle,|1\rangle\}$, is found to be,
\begin{small}
\begin{equation}\label{NUU}
\begin{aligned} U&=\left[\begin{array}{cc}e^{-i\frac{\eta_{-}}{2}}\left(X_{2\gamma,\chi_{-}}+iY_{2\gamma,\chi_{+}}\right)& e^{-i\frac{\eta_{+}}{2}}\left(iZ_{2\gamma,\chi_{+}}-Y_{\chi_{-},2\gamma}\right) \\ e^{i\frac{\eta_{+}}{2}}\left(iZ_{2\gamma,\chi_{+}}+Y_{\chi_{-},2\gamma}\right) & e^{i\frac{\eta_{-}}{2}}\left(X_{2\gamma,\chi_{-}}-iY_{2\gamma,\chi_{+}}\right)\end{array}\right] \end{aligned},
\end{equation}
\end{small}
where  $X_{a,b}\equiv\cos\frac{a}{2}\cos\frac{b}{2}$, $Y_{a,b}\equiv\sin\frac{a}{2}\cos\frac{b}{2}$, $Z_{a,b}\equiv\sin\frac{a}{2}\sin\frac{b}{2}$, $\chi_{\pm}=\chi(\tau)\pm\chi(0)$ and $\eta_{\pm}=\eta(\tau)\pm\eta(0)$. The initial values of the auxiliary variables $\eta$ and $\chi$ can be determined by the target geometric gates.

\begin{figure}[tbp]
	\centering
\includegraphics[width=8.5cm]{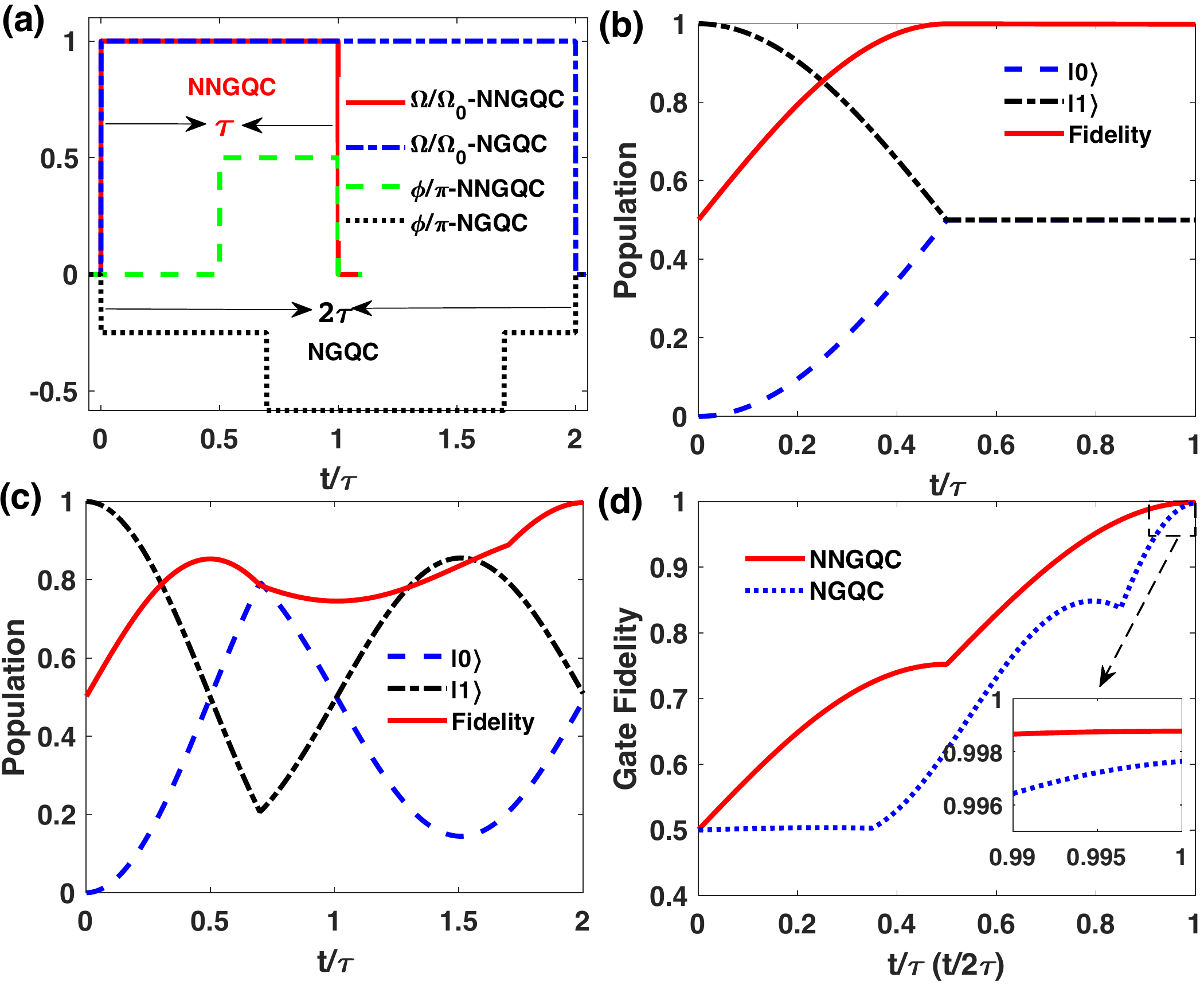}
\caption{\label{fig2} (a) The Rabi frequency $\Omega(t)$ and phase $\varphi(t)$ of $U_{1}$ gate for NNGQC and NGQC. State population and state fidelity of (b) NNGQC and (c)~NGQC with the initial state being $|0\rangle$. (d) Gate fidelity of $U_{1}$ as a function of $t/\tau$ ($t/2\tau$).}
\end{figure}

\begin{figure*}[htbp]
	\centering
\includegraphics[width=15cm]{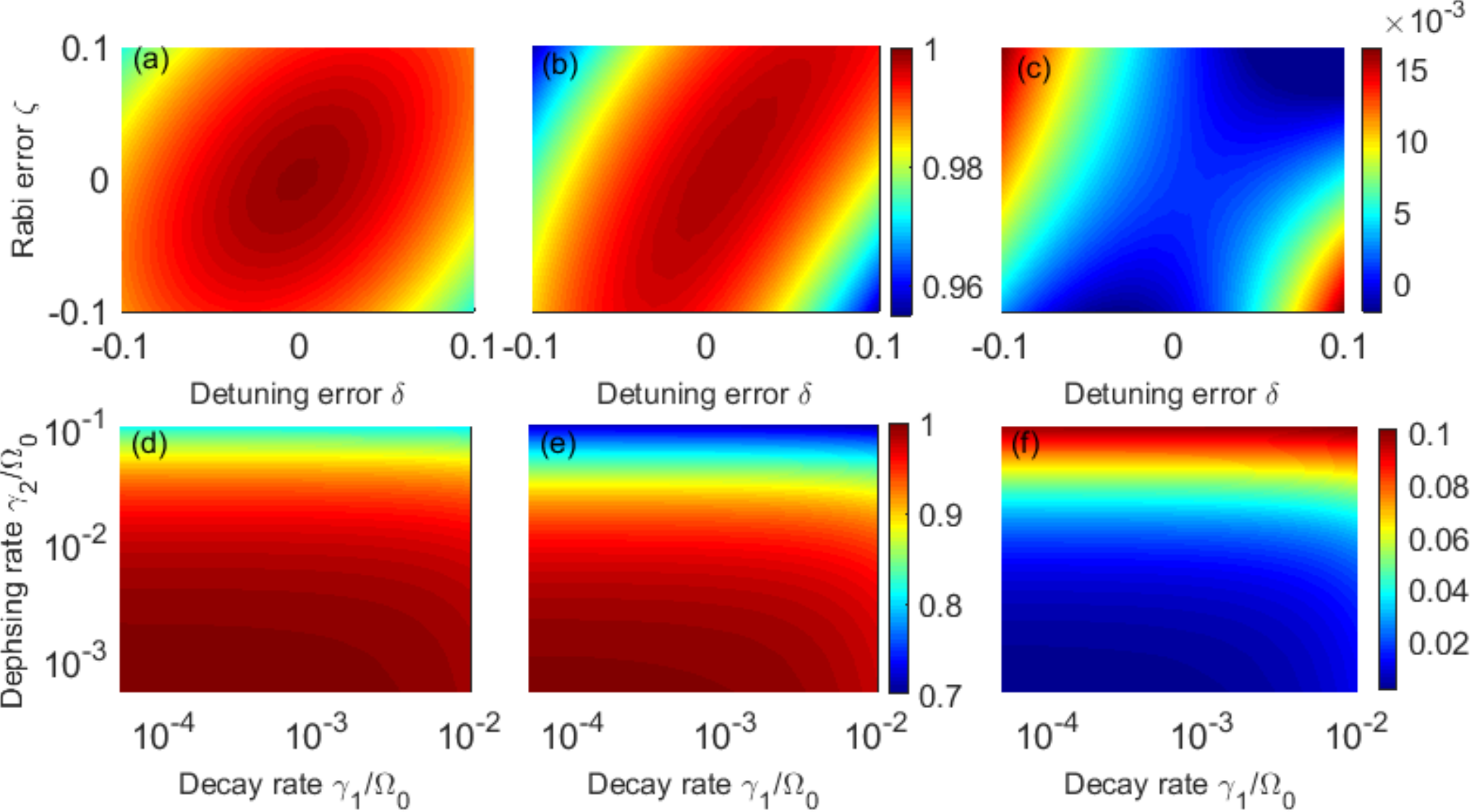}
\caption{\label{fig3} The performance of $U_{1}$ gate under imperfections. Gate fidelities of (a) NNGQC, (b) NGQC under the Rabi error $\zeta$ and detuning error $\delta$. (c) is the difference between (a) and (b).  The gate fidelities for~(d) NNGQC, (e) NGQC and (f) difference as a function of decay rate $\gamma_{1}$ and dephasing rate $\gamma_{2}$, respectively.}
\end{figure*}

To illustrate the geometric rotation, $U$ is conveniently represented (ignore a global phase) as,
\begin{equation}\label{rotation}
U(\theta, \alpha, \beta)=Z_{\beta}X_{\theta}Z_{\alpha}
\end{equation}
where $\theta \equiv \sin ^{-1}(\sqrt{Z_{2\gamma, \chi_{+}}^{2}+Y_{\chi_{-}, 2\gamma}^{2}})$, $\alpha\equiv-\tan ^{-1}\left(\frac{Y_{2\gamma, \chi_{+}}}{X_{2\gamma, \chi_{-}}}\right)-\tan ^{-1}\left(\frac{Z_{2\gamma, \chi_{+}}}{Y_{\chi_{-},2\gamma}}\right)+\frac{\eta_{-}-\eta_{+}-\pi}{2}$ and $\beta \equiv-\tan ^{-1}\left(\frac{Y_{2\gamma, \chi_{+}}}{X_{2\gamma, \chi_{-}}}\right) +\tan ^{-1}\left(\frac{Z_{2\gamma,\chi_{+}}}{Y_{\chi_{-},2\gamma}}\right)+\frac{\eta_{-}+\eta_{+}+\pi}{2}$ are rotation of angles around the $X$ and $Z$ axis of the Bloch sphere, respectively. Any single-qubit $SU(2)$ operation can be realized with $U(\theta, \alpha, \beta)$ by choosing the geometric phase $\gamma$, the initial and final value $\eta_{\pm}$ and $\chi_{\pm}$. For example, we can realize the noncyclic geometric $U_{1}=U(\pi/2,-\pi/2,0)$ and $U_{2}=U(\pi/2,\pi/2,\pi/2)$  (Hardmard gate) by setting the parameters as  $\{\gamma=\frac{\pi}{4}, \chi_{+}=0, \chi_{-}=\pi,\eta_{\pm}=\mp\frac{\pi}{2}\}$, and $\{\gamma=\frac{\pi}{4}, \chi_{+}=-\frac{3\pi}{2}, \chi_{-}=\frac{\pi}{2},\eta_{\pm}=\mp\frac{\pi}{2}\}$, respectively. More specifically, we note that Eq. (\ref{NUU}) is the $Z$-rotation gate $Z_{(-\eta_{-})}=\exp(i\frac{\eta_{-}}{2}Z)$ when $\gamma=\pi,\chi_{-}=0$ and is the $X$-rotation gate $X_{\chi_{-}}=\exp(i\frac{\chi_{-}}{2}X)$ when $\gamma=\pi$, $\eta_{-}=0$ and $\eta_{+}=-\pi$.

To further understand the scheme of our NNGQC, we found that the non-diagonal parts of ${\bf A}$ and ${\bf K}$ satisfy the relations of unconventional quantum holonomy~\cite{zhu2003,liu2020} as $\int_{0}^{\tau}{\bf A}_{km} dt/\int_{0}^{\tau}{\bf K}_{km} dt=-1$ for $k \neq m$, where ${\bf A}_{km}(t)=\frac{1}{2}\Omega\left[\cos\chi\cos(\varphi-\eta)-i\sin(\varphi-\eta)\right]$. Although ${\bf A}+{\bf K}$ is a diagonal matrix, ${\bf A}$ and ${\bf K}$ are both non-diagonal in our scheme. Specifically, ${\bf A}$ does represent a non-abelian connection with non-vanishing commutation relation $[{\bf A}(t),{\bf A}(t')]\neq0$, which proves the non-Abelian nature of the gate in Eq. (\ref{NUU})~\cite{b4}. In addition, we emphasize that the auxiliary states $|\phi_{1,2}(t)\rangle$ in our model are chosen generally for the resonate two-level system and thus the choice of auxiliary states does not change the above results.

Now, to construct the NNGQC gate, one simple parameter set of choice is found to be $\chi(t)=\Omega_{0}t-\chi_{0}, \quad \eta(t)=\phi_{1}\epsilon(t)+\phi_{0}$, where the step function $\epsilon(t)$ satisfies $\epsilon=0$ with $t\in[0,\frac{\chi_{0}}{\Omega_{0}}]$ and $\epsilon=1$ with $t\in[\frac{\chi_{0}}{\Omega_{0}},\tau]$ and $\Omega_{0}$, $\phi_{1}$, $\phi_{0}$ and $\chi_{0}$ are constants. With the settings, we can obtain  the initial and final value $\eta_{\pm}$ and $\chi_{\pm}$ as $\eta_{+}=\phi_{1}+2\phi_{0}$, $\eta_{-}=\phi_{1}$, $\chi_{+}=\Omega_{0}\tau-2\chi_{0}$, $\chi_{-}=\Omega_{0}\tau$. Meanwhile, the geometirc phase is taken by $\gamma=\int_{0}^{\tau}\frac{1}{2}\dot{\eta}\cos\chi dt=\frac{\phi_{1}}{2}$. For $U_{1}$ gate, the control parameters are chosen as $\chi_{0}=\pi/2$, $\phi_{0}=-\pi/2$, $\phi_{1}=\pi/2$, and $\tau=\frac{\pi}{\Omega_{0}}$.In sharp contrast to the conventional NGQC schemes to construct $U_{1}$~\cite{zhao2017,Chen2018,Zhang2020,XXu2018,ZZhao2019} that requires evolution time being $2\tau$, by choosing the same maximum Rabi frquecy, we found that the required gate time of NNGQC to construct $U_{1}$ is $\tau$ [as shown in Fig. \ref{fig2}(a)], which is reduced by 50\% (see Appendix A for details).

\section{Gate performance and robustness}

The performance of the $U_{1}$ gate can be simulated by using the Lindblad master equation~\cite{Lindblad}.
In our simulation, we have used the following set of experimental parameter~\cite{Olmschenk2010,xia2015L,Wang2015}. The Rabi frequency, decay and dephasing rates are set as $\Omega_{0}=2\pi \times 6.25$ kHz,  $\gamma_{1}\approx2\Omega_{0}\times10^{-4}$ and $\gamma_{2} \approx 2\Omega_{0}\times10^{-3}$ corresponding to $T_{1}=590$ms and $T_{2}=50$ms.  
Suppose that the qubit is initially prepared in the $|\psi(0)\rangle=|0\rangle$ state, the time-dependence of the state populations and the state fidelity $F=\left|\left\langle\psi_{I} | \psi(\tau)\right\rangle\right|^{2}$ of realizing the $U_{1}$ gate for NNGQC and NGQC are depicted in Fig. \ref{fig2}(b) and \ref{fig2}(c), where the state fidelities of NNGQC and NGQC are obtained to be 99.87\% and 99.75\%, respectively. Furthermore, we have also investigated the gate fidelity of $U_{1}$ defined by $F=(1 / 2 \pi) \int_{0}^{2 \pi}\left\langle\psi_{I}\left|\rho\right| \psi_{I}\right\rangle d \Theta$ for initial states of the form $|\psi\rangle=\cos\Theta|0\rangle+\sin\Theta|1\rangle$, where a total of 1001 different values of $\Theta$ were uniformly chosen in the range of $[0, 2\pi]$, as shown in Fig. \ref{fig2}(d). We found that the gate error $(1-F)$ of NNGQC can be reduced by as much as 50\% compared with the gate error of NGQC (0.24\%).

Now, we start to demonstrate the robustness of our scheme.  We firstly consider the robustness of our NNGQC against Rabi errors and assume the amplitudes of control pulse to vary in the range of  $\Omega_{0}\rightarrow(1 + \zeta) {\Omega _{0}}$ with the error fraction $\zeta\in[-0.1,0.1]$. Next, we take the detuning noise to be $\Delta\sigma_{z}$ with $\Delta=\delta\Omega_{0}$ being static and the fraction is $\delta\in[-0.1,0.1]$. Comparing our NNGQC with the conventional NGQC methods, we plot the performance of the same geometric gate with the same pulse error. As shown in Fig.~\ref{fig3}(a), Fig.~\ref{fig3}(b) and Fig.~\ref{fig3}(c), the NNGQC is always more robust than the NGQC gate. Furthermore,
we also simulated the gate fidelity as a function of decay rate and dephasing rate $\gamma_{1}$ and $\gamma_{2}$. For above two schemes as shown in Fig.~\ref{fig3}(d), Fig.~\ref{fig3}(e) and  Fig.~\ref{fig3}(f), our scheme of NNGQC can greatly suppress the decoherence effect comparing with the conventional NGQC.

To further show the noise-resilient geometric feature of our NNGQC, we also take the above noises to compare the performance of our geometric gate with that of the corresponding dynamical gate (DG)~\cite{Zheng2016,Barends2014} as shown in Appendix B. From the numerical result, we can clearly see that our NNGQC scheme is more robust against both the pulse control error and decoherence error than DG scheme.

\section{Nontrivial two-qubit gate}
\begin{figure}[bp]
	\centering
\includegraphics[width=1.0\linewidth]{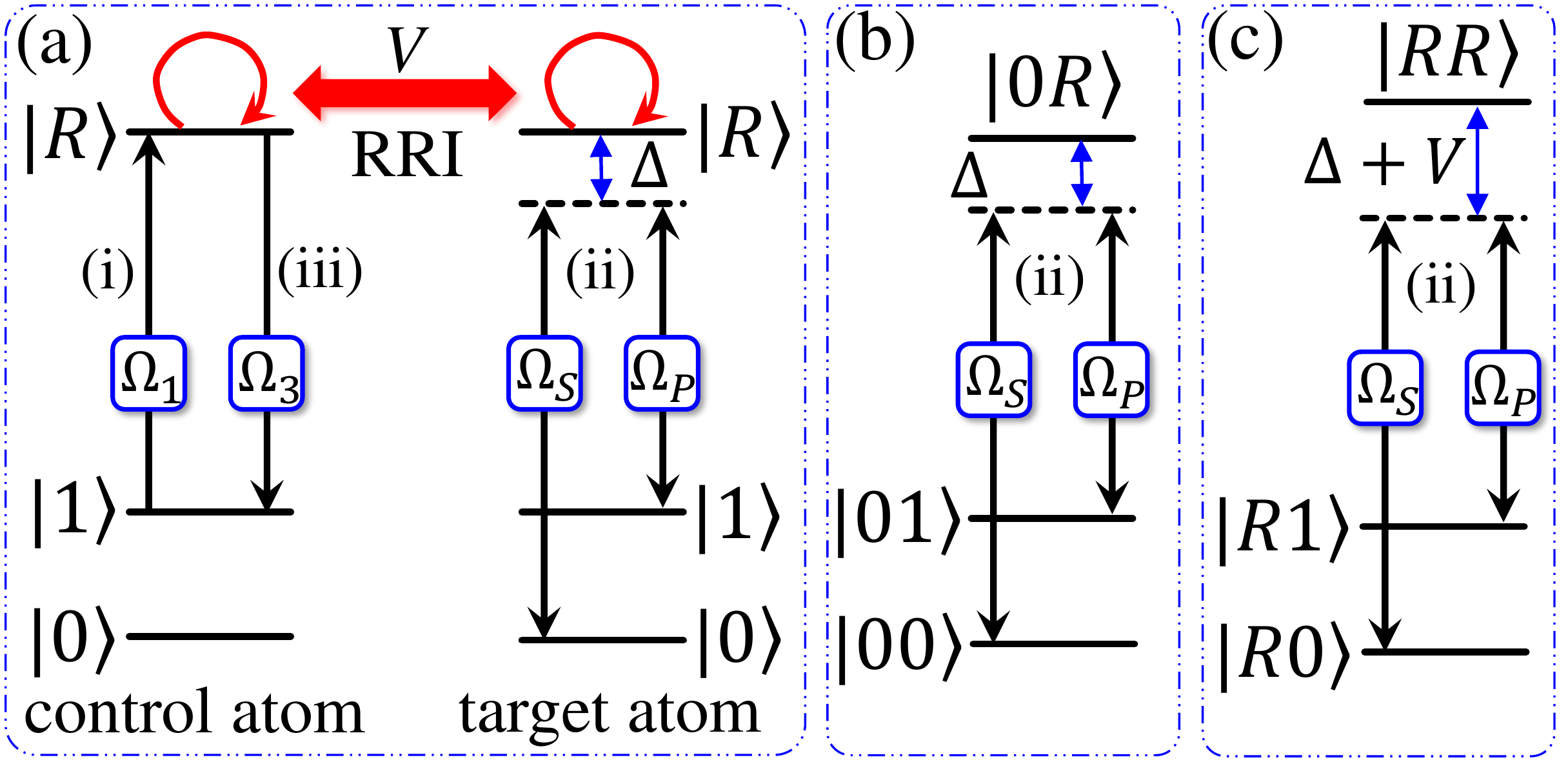}
\caption{\label{fig4} (a) Illustration of the two-qubit NNGQC gate based on unconventional Rydberg blockade with single-atom basis. \emph{V} denotes the RRI strength. For control atom, $|1\rangle$ is coupled with $|R\rangle$ through two-photon process with Rabi frequency $\Omega_{1}(\Omega_{3})$ in step (i)[(iii)]. For target atom, $|0\rangle~(|1\rangle)$ is coupled with $|R\rangle$ with Rabi frequency $\Omega_{S}~(\Omega_{P})$ and detuning $\Delta$ via two-photon process in step~(ii). $\Omega_{j} = |\Omega_{j}|e^{i\varphi_{j}}$ with $j = 1,~3,~S,~P$. (b)[(c)] Dynamical process of step~(ii) under two-atom basis in the without~(with)~RRI. For conventional blockade, the dynamical process in (b) is resonant while in (c) is detuned by \emph{V} and always be discarded. However, in our scheme both of the dynamical processes in (b) and (c) are used for the gate, and thus one can call it "unconventional Rydberg blockade".}
\end{figure}

In this section, we proceed to implement nontrivial two-qubit Rydberg quantum gates free from blockade error with the pulse similar to that designed in single-qubit case. As shown in Fig.~\ref{fig4}, we consider two $^{133}$Cs atoms with magnetically insensitive ``clock'' states encoding $|0\rangle\equiv|6S_{1/2},~F = 3,~ m_{F}=0\rangle$ and $|1\rangle\equiv|6S_{1/2},~F = 4,~ m_{F}=0\rangle$~\cite{Jaunatphys}. The Rydberg state is chosen as $|R\rangle \equiv |61S_{1/2}\rangle$. And the $C_{6}$ parameter can be evaluated as 126~GHz$\cdot\mu m^6$~\cite{Rydberginteraction}. 

The basic process of two-qubit gate is shown in Fig.~\ref{fig4}, where the control in resonant interaction but the target atom is large-detuning interaction. The required three steps are as follows.

\emph{Step~(i)}. Turn on the laser on control atom with Hamiltonian 
\begin{equation}
    H_{c} = \frac{\Omega_{1}(t)}{2}|1\rangle\langle R| + \rm{H.c.},
\end{equation}
where $\Omega_{1}(t)\equiv|\Omega_{1}(t)|e^{i\varphi_{1}(t)}$. We set $\varphi_{1}(t)=0$ and $\int\Omega_{1}(t)dt=\pi$ in step~(i).

\emph{Step~(ii)}. Turn off the laser on control atom and turn on lasers with Rabi frequencies $\Omega_{S}$ and $\Omega_{P}$ on target atom. The Hamiltonian is 
\begin{equation}
    H_{t} = \Delta|R\rangle\langle R|+\frac{1}{2}\Big[\Omega_{S}(t)|0\rangle + \Omega_{P}(t)|1\rangle\Big]\langle R| + {\rm H.c.},
\end{equation}
where $\Omega_{S}(t)\equiv|\Omega_{S}(t)|e^{i\varphi_{S}(t)}$ and $\Omega_{P}(t)\equiv|\Omega_{P}(t)|e^{i\varphi_{P}(t)}$.
Besides, the RRI Hamiltonian 
\begin{equation}
    H_{V} = V|RR\rangle\langle RR|
\end{equation}
may be in existence conditioned on the control atom is excited or not. We use $|mn\rangle\langle mn|$ to denote the abbreviation of $|m\rangle\langle m|\otimes|n\rangle\langle n|$ here and throughout the manuscript for simplify. 
Thus, the dynamical process can be classified as two cases in Fig.~\ref{fig4}(b) and Fig.~\ref{fig4}(c), respectively, depend whether the control atom is not excited or excited. 

\begin{figure}[htbp]
	\centering
\includegraphics[width=8.5cm]{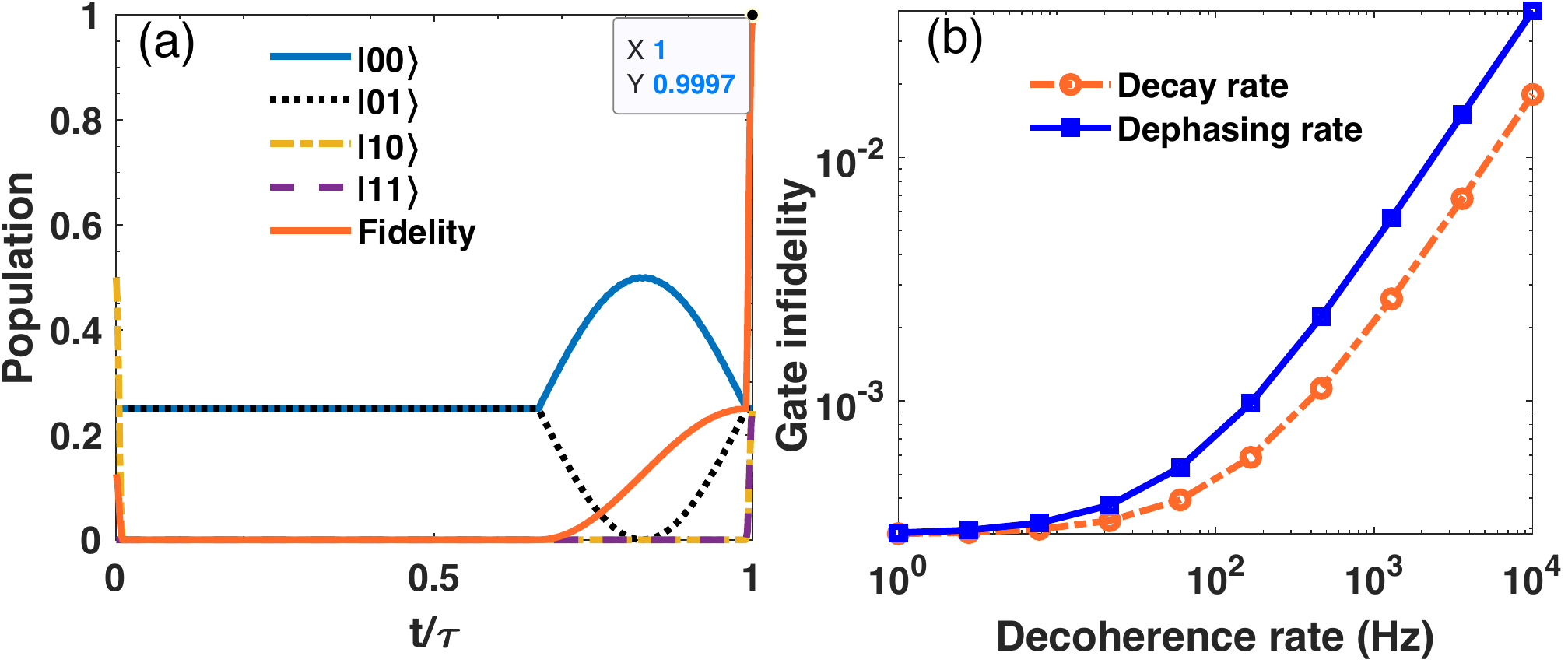}
\caption{\label{fig5} (a) State population and state fidelity of two-qubit gate with the initial state being $\frac{1}{2}(|00\rangle+|01\rangle+\sqrt{2}|10\rangle)$. (b) Gate infidelity as a function of decay rate and dephasing rate for Rydberg state .}
\end{figure}

The effective Hamiltonian in Fig.~\ref{fig4}(b) can be calculated as
\begin{equation}\label{Hamiltonian4b}
    H_{\rm eff,1} = \frac{\Omega_{\rm eff,1}}{2}|00\rangle\langle01| + {\rm H.c.},
\end{equation}
where $\Omega_{\rm eff,1}=\Omega_{S}\Omega_{P}^*/(2\Delta)$ and the stark shifts are vanished when $|\Omega_{S}| = |\Omega_{P}|$.
Similarly, the effective Hamiltonian in Fig.~\ref{fig4}(c) would be 
\begin{equation}\label{Hamiltonian4c}
    H_{\rm eff,2} = \frac{\Omega_{\rm eff,2}}{2}|R0\rangle\langle R1| + {\rm H.c.},
\end{equation}
where $\Omega_{\rm eff,2}=\Omega_{S}\Omega_{P}^*/[2(\Delta+V)]$ and the stark shifts are also vanished when $|\Omega_{S}| = |\Omega_{P}|$. 

Equations~(\ref{Hamiltonian4b}) and (\ref{Hamiltonian4c}) have similar form as Eq.~(\ref{Original}). Thus, we can use the similar pulses to construct the noncyclic nonadiabatic geometric operations. That is, in step~(ii), one can get the operation
\begin{equation}
    \mathcal{U}_{2} = |0\rangle_{c}\langle0|\otimes U(\theta_{1},\alpha_{1},\beta_{1}) + |R\rangle_{c}\langle R|\otimes U(\theta_{2},\alpha_{2},\beta_{2})
\end{equation}
with suitable laser parameters.

\emph{Step(iii)}. Turn off the lasers on target atom and at the same time turn on laser with Rabi frequency $\Omega_{3}(t)$ on control atom. If $|\Omega_{3}(t-t_{2}-t_{1})| = |\Omega_{1}(t)|$ and $\varphi_{3}=\pi$, in which $t_{1(2)}$ denotes the evolution time in step~(i)[(ii)], one can get the the whole evolution operator as 
\begin{equation}\label{twqo}
    \mathcal{U} = |0\rangle_{c}\langle0|\otimes U(\theta_{1},\alpha_{1},\beta_{1}) + |1\rangle_{c}\langle 1|\otimes U(\theta_{2},\alpha_{2},\beta_{2}).
\end{equation}
Therefore, in general, we know that  Eq. (\ref{twqo}) represents a nontrivial two-qubit entangled gate, since $U$ in subspace $\{|00\rangle,|01\rangle\}$ and $\{|10\rangle,|11\rangle\}$ is different. 

When $\theta_{1}=\pi/2,\alpha_{1}=\pi/2,\beta_{1}=\pi/2$ and $\Delta=V$, we obtain two-qubit entangled gate with matrix representation as
\begin{equation}\label{UTW}
\mathcal{U}=\left(\begin{array}{cccc}0& i & 0 & 0 \\ -i& 0 & 0 & 0 \\ 0 & 0 & \frac{e^{i\pi/4}}{\sqrt{2}}& \frac{e^{i\pi/4}}{\sqrt{2}} \\ 0 & 0 & \frac{e^{i\pi/4}}{\sqrt{2}}& \frac{-e^{i\pi/4}}{\sqrt{2}}\end{array}\right).
\end{equation}
Therefore, our scheme is sufficient for universal quantum computation when assisted by a combination of the single-and two-qubit gate. 

To evaluate the performance of two-qubit entangled gate, we take the parameters from the state-of-art experiments as the Rabi frequency $\Omega_{0}=\Omega_{1}=2\pi\times10$~MHz and the detuning $\Delta\approx17\Omega_{0}$, where $\Omega_{P}$ and $\Omega_{S}$ are governed by the Eq. (2) in maintext. As shown in Fig.\ref{fig5}(a), we plot the state populations and the state fidelity of the two-qubit gate with the initial state $|\psi(0)\rangle=\frac{1}{2}(|00\rangle+|01\rangle+\sqrt{2}|10\rangle)$, where the state fidelity is obtained to be 99.97\% without considering relaxation. Moreover, we have also investigated the gate infidelity~\cite{Nielsen2002,White2007} of two-qubit $1-F$  as a function of decay rate and dephasing rate of Rydberg state as shown in Fig. \ref{fig5}(b), and found that our two-qubit geometric gate is robust against decoherence from the environmental noises.

\section{Conclusion}
In summary, we have presented a new framework of NNGQC, which universal nonadiabatic geometric gates can be constructed via noncyclic non-Abelian geometric phase.  Comparing with conventional NGQC, NNGQC can further reduce the geometric gate time beyond the limitation of cyclic condition. Consequently, our proposal is more robust against the decay and dephasing effects from the environmental decoherence. Moreover, we construct a nontrivial two-qubit geometric gate via RRI-induced large detuning process seriously without discarding the process induced by RRI-induced ``blockade'' terms. Therefore, our scheme provides a promising way towards fault-tolerant quantum computation for neutral-atom-based quantum system.

\acknowledgments 
This work is supported by the Key-Area Research and Development Program of Guangdong Province (Grant No.2018B030326001), the National Natural Science Foundation of China (Grant No.11875160, No.11874156 and No.11804308), the Natural Science Foundation of Guangdong Province (Grant No.2017B030308003), the National Key R\& D Program of China (Grant No.2016YFA0301803, the Guangdong Innovative and Entrepreneurial Research Team Program (Grant No.2016ZT06D348), the Economy, Trade and Information Commission of Shenzhen Municipality (Grant No.201901161512), the Science, Technology and Innovation Commission of Shenzhen Municipality (Grant No. JCYJ20170412152620376, No. JCYJ20170817105046702, and No. KYTDPT20181011104202253)).

\appendix
\section{ Nonadiabatic geometirc quantum computation}
Here, we shall succinctly derive NGQC~\cite{zhao2017,Chen2018,Zhang2020,XXu2018,ZZhao2019} in our framework to provide a unified view on geometric quantu computation. For conventional NGQC, it should satisfy the cyclic condition as $\left|\phi_{1}(\tau)\right\rangle=\left|\phi_{1}(0)\right\rangle$. Consequently, the Rabi frequency is governed by the following equation 
\begin{equation}\label{cyclic}
\int_{0}^{T} \Omega(t) d t=2\pi \ ,
\end{equation}
where $T$ denotes the gate time.
In this way, the cyclic nonadiabatic geometric gate becomes: 
\begin{equation}\label{cyclic}
\begin{aligned} U(\tau,0) &=e^{i\gamma}\left|\phi_{1}(0)\right\rangle\langle\phi_{1}(0)|+e^{-i\gamma}| \phi_{2}(0)\rangle\left\langle\phi_{2}(0)\right| \\ &=\cos \gamma+i \sin \gamma\left(\begin{array}{cc}\cos \mu & \sin \mu e^{-i \eta_{0}} \\ \sin \mu e^{i \eta_{0}} & -\cos \mu\end{array}\right) \\ &=e^{i \gamma \vec{n} \cdot \vec{\sigma}} \end{aligned} \ ,
\end{equation}
where $\mu=\chi(0)$ and $\eta_{0}=\eta(0)$ denotes the initial value of $\chi$ and $\eta$ at time $t=0$; $\vec{n}=(\sin \mu\cos \eta_{0}, \sin \mu \sin \eta_{0}, \cos \mu)$, and $\vec{\sigma}$ are the Pauli matrices. 

\begin{figure}[htbp]
	\centering
\includegraphics[width=8.5cm]{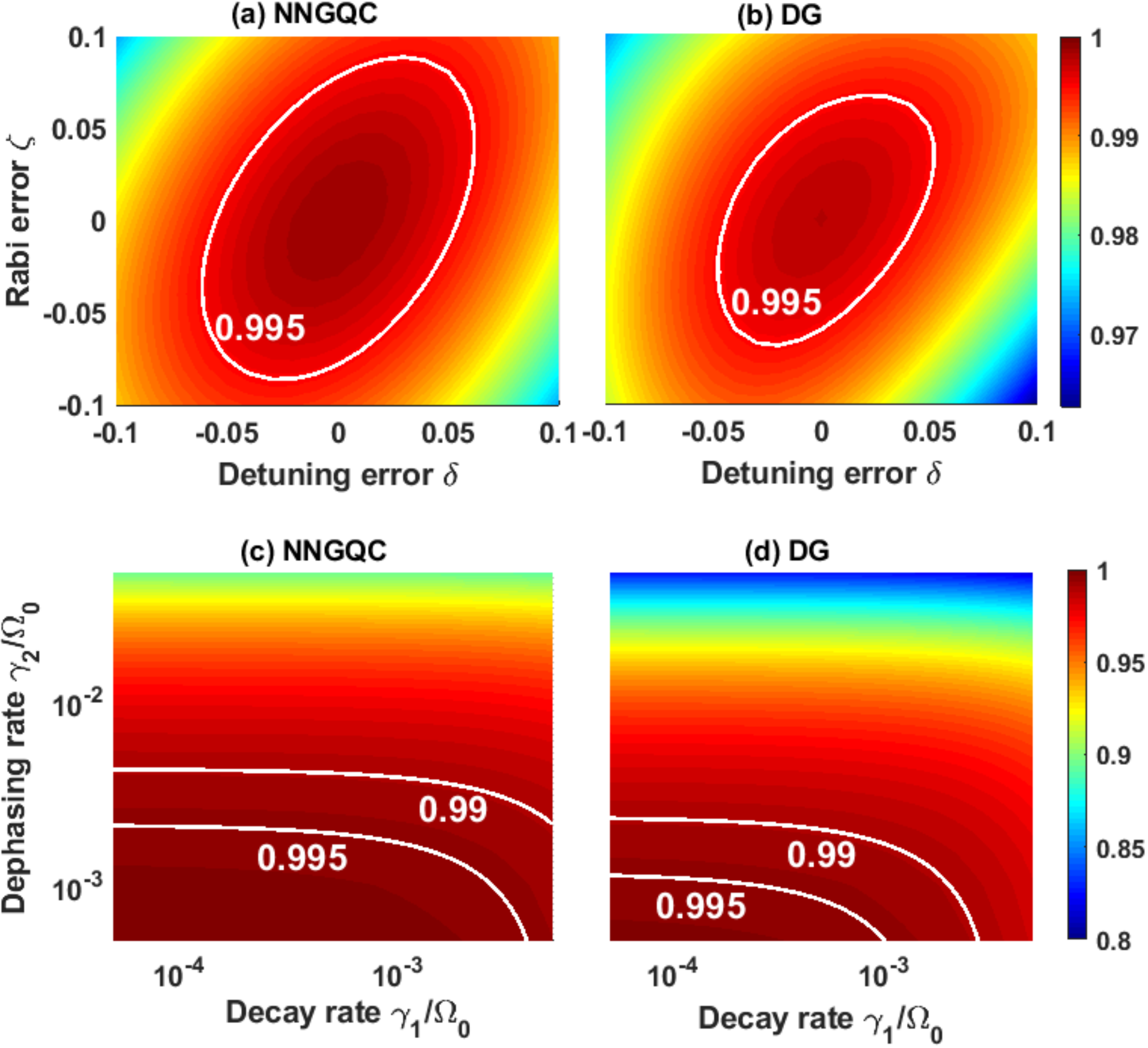}
\caption{\label{fig6} Gate fidelities of (a) NNGQC, and (b) DG under the Rabi error $\zeta$ and detuning error $\delta$.  The gate fidelities for~(c) NNGQC, and (d) DG as a function of decay rate $\gamma_{1}$ and dephasing rate $\gamma_{2}$, respectively.}
\end{figure}

To realize the $U_{1}$ gate of NGQC, we take parameters as $\gamma=3\pi/4$, $\eta_{0}=\pi/4$, and $\mu=7\pi/20$ corresponding the Rabi frequency $\Omega(t)=\Omega_{0}$ and the evolution time $T=2\pi/\Omega_{0}=2\tau$, as shown in Fig. 2(a).

\section{Robustness of NNGQC vs dynamical gate}

Before comparing the robustness of our NNGQC gate with the dynamical gate (DG)~\cite{Zheng2016,Barends2014}, let ’s briefly talk about how to realize the $U_{1}$ dynamical gate. When this Hamiltonian $H$ is time-independent, we can obtain the evolution operator as
\begin{equation}\label{DGgate}
U_{dg}(\Omega\tau_{1},\varphi)=\left(\begin{array}{cc}\cos (\frac{\Omega\tau_{1}}{2}) & -i e^{-i \varphi} \sin (\frac{\Omega\tau_{1}}{2}) \\ -i e^{i \varphi} \sin (\frac{\Omega\tau_{1}}{2}) & \cos (\frac{\Omega\tau_{1}}{2})\end{array}\right) \ .
\end{equation} 

In general, any single-qubit gate can be realized by the evolution operator. 
We note that Eq. (\ref{DGgate}) is X-rotation  $U_{dg}=X_{\Omega\tau_{1}}$ when $\varphi=0$. To realize a Z-rotation gate $U_{dg}=Z_{\phi}$, we need to take two sequential evolutions $U_{dg}(\pi,\phi/2)$ and $U_{dg}(\pi,-\phi/2)$. In this way, the general dyamical gate is given by 
\begin{equation}\label{DGG}
U_{dg}=Z_{\phi}X_{\Omega\tau_{1}}Z_{\alpha} \ .
\end{equation}
For the $U_{1}$ dynamical gate, we take parameters as $\Omega(t)=\Omega_{0}$, $\tau_{1}=\pi/2\Omega_{0}=\frac{1}{2}\tau$, $\phi=0$, $\alpha=-\pi/2$, and $T=5\tau/2$.

Comparison of the robustness against the detuning error and Rabi frequency, of all the two implementations are shown in Fig. \ref{fig6} (a) and \ref{fig6} (b). From the numerical result, we can clearly see that the NNGQC scheme is more robust against the pulse control errors than the corresponding DG scheme. Furthermore, we also consider comparison of the robustness against the decay rate and dephasing rate caused by environmental noise. From the Fig. \ref{fig6} (c) and \ref{fig6} (d), we know that our scheme of NNGQC can suppress the decoherence effect comparing with the DG.


\end{document}